\documentclass{kapedbk} 

\usepackage[dvips]{graphicx}

\upperandlowercase

\kluwerbib 

\newcommand{\araa}{Ann. Rev. of Astr. \& Ap.}
\newcommand{\apj}{Astrophys. J.}
\newcommand{\apjl}{Astrophys. J., Lett.}
\newcommand{\apjs}{Astrophys. J., Supp.}
\newcommand{\aap}{Astr. \& Ap.}
\newcommand{\baas}{Bull. Am. Astr. Soc.}
\newcommand{\nat}{Nature}
\newcommand{\prl}{Phys. Rev. Lett.}
\newcommand{\prd}{Phys. Rev. D}
\newcommand{\mnras}{Mon. Not. R. A. S.}
\newcommand{\iMAP}{{\em WMAP}}

\def\plotone#1{\centering \leavevmode
 \includegraphics[width=.95\columnwidth]{#1}}

\def\plottwo#1#2{\centering \leavevmode
\includegraphics[width=.45\columnwidth]{#1} \hfil
\includegraphics[width=.45\columnwidth]{#2}}

\begin{document}
\articletitle[CMB Observational Techniques and Recent Results]
{CMB Observational Techniques \\
and Recent Results
}

\author{Edward L. Wright\altaffilmark{1}}

\altaffiltext{1}{UCLA Astronomy\\
PO Box 951562\\Los Angeles, CA 90095-1562\\USA}
\email{wright@astro.ucla.edu}

\begin{abstract}
The Cosmic Microwave Background (CMB) consists of photons that were last
created about 2 months after the Big Bang, and last scattered about
380,000 years after the Big Bang.  The spectrum of the CMB is very close
to a blackbody at $2.725\;\mbox{K}$, 
and upper limits on any deviations from of 
the CMB from a blackbody place strong constraints on energy transfer
between the CMB and matter at all redshifts less than 2 million.
The CMB is very nearly isotropic, but a dipole anisotropy of
$\pm 3.346(17)\;\mbox{mK}$ shows that the Solar System barycenter
is moving at $368 \pm 2\;\mbox{km/sec}$ relative to the observable
Universe.
The dipole corresponds to a spherical harmonic index $\ell=1$.  The
higher indices $\ell \geq 2$ indicate intrinsic inhomogeneities in
the Universe that existed at the time of last scattering.
While the photons have traveled freely only since the time of
last scattering, the inhomogeneities traced by the CMB photons
have been in place since the inflationary epoch only $10^{-35}\;\mbox{sec}$
after the Big Bang.
These intrinsic anisotropies are much smaller in amplitude than the
dipole anisotropy, with $\Delta T \leq 100\;\mbox{$\mu$K}$.
Electron scattering of the anisotropic radiation field produces
an anisotropic linear polarization in the CMB with amplitudes
$\leq 5\;\mbox{$\mu$K}$.
Detailed studies of the angular power spectrum of the
temperature and linear polarization anisotropies have yielded 
precise values for many cosmological parameters.
This paper will discuss the techniques necessary to measure signals
that are 100 million times smaller than the emission from the instrument
and briefly describe results from experiments up to \iMAP.
\end{abstract}


\section{Introduction}

The Cosmic Microwave Background (CMB) was first seen via its effect
on the interstellar CN radical (\cite{adams:1941}) but the
significance of the this datum was not realized until after 1965
(\cite{thaddeus:1972}; \cite{kaiser/wright:1990}).
In fact, \cite{herzberg:1950}\ calculated a $2.3\;\mbox{K}$ excitation
temperature for the CN transition and said it had ``of course only a very
restricted meaning.''
Later work by
\cite{roth/meyer/hawkins:1993} obtained a value for $T_\circ =
2.729^{+0.023}_{-0.031}\;\mbox{K}$ at the CN 1-0 wavelength of
$2.64\;\mbox{mm}$ which is still remarkably accurate.

\begin{figure}[ht]
\plotone{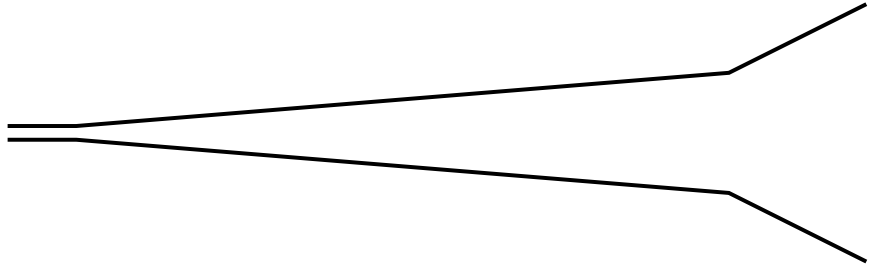}
\caption{The flared horn used by Dicke during World War II.
\label{fig:Dicke-horn}}
\end{figure}

A second notable missed opportunity to discover the CMB occurred
during World War II radar research.  In a paper reporting 
measurements of the atmospheric opacity at $1.25\;\mbox{cm}$
wavelength using zenith angle scans, an upper limit of
$T_\circ < 20\;\mbox{K}$ is given (\cite{dicke/etal:1946}).
The Dicke switch and the differential radiometer were invented
for this work.  Since the reference load was at room temperature,
the large difference signal of $250\;\mbox{K}$ at the zenith did
not allow for a precise determination of the cosmic temperature.  
The antenna was a flared horn (Figure \ref{fig:Dicke-horn}) which
was specifically designed for low sidelobes.
This missed opportunity is especially ironic since Dicke was 
actually building a radiometer to look for the CMB when
he heard about the \cite{penzias/wilson:1965}
result (\cite{dicke/etal:1965}).

\begin{figure}[ht]
\plotone{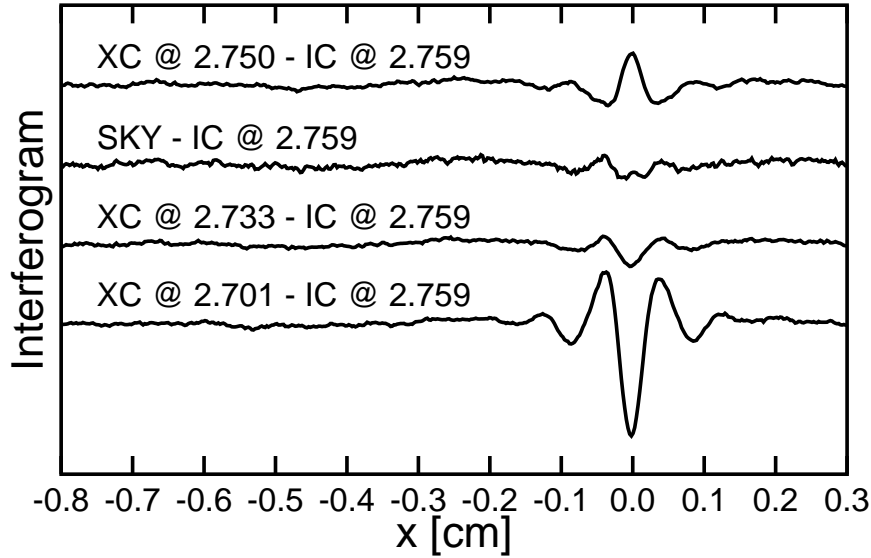}
\caption{Four interferograms taken by FIRAS.  Three show observations
of the external calibrator at different temperatures straddling the
sky temperature, while the fourth shows observations of the sky.
All were taken with the internal calibrator at 2.759 K.\label{fig:ifg4}}
\end{figure}

\begin{figure}[ht]
\plotone{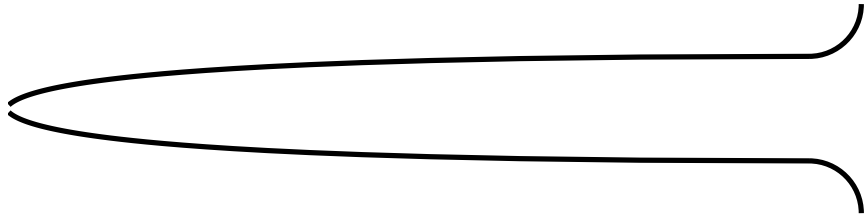}
\caption{The FIRAS horn: a compound parabolic concentrator with
a trumpet bell flare to reduce sidelobes.\label{fig:FIRAS-horn}}
\end{figure}

The most accurate measurements of the CMB spectrum to date have come
from the Far InfraRed Absolute Spectrophotometer (FIRAS) on the
COsmic Background Explorer (COBE) (\cite{boggess/etal:1992}).
In contradiction to its name, FIRAS was a fully differential
spectrograph that only measured the difference between the sky and
an internal reference source that was very nearly a blackbody.
Figure \ref{fig:ifg4} shows the interferograms observed by FIRAS
for the sky and for the external calibrator (XC) at three
different temperatures, all taken with the internal calibrator (IC)
at $2.759\;\mbox{K}$.
Data from the entire FIRAS dataset show that the rms deviation
from a blackbody is only 50 parts per million of the peak $I_\nu$
of the blackbody (\cite{fixsen/etal:1996}) and a recalibration
of the thermometers on the external calibrator yield a blackbody
temperature of $2.725 \pm 0.001\;\mbox{K}$ (\cite{mather/etal:1999}).
FIRAS also had a flared horn to reduce sidelobes as seen in
Figure \ref{fig:FIRAS-horn}.

\begin{figure}[ht]
\plotone{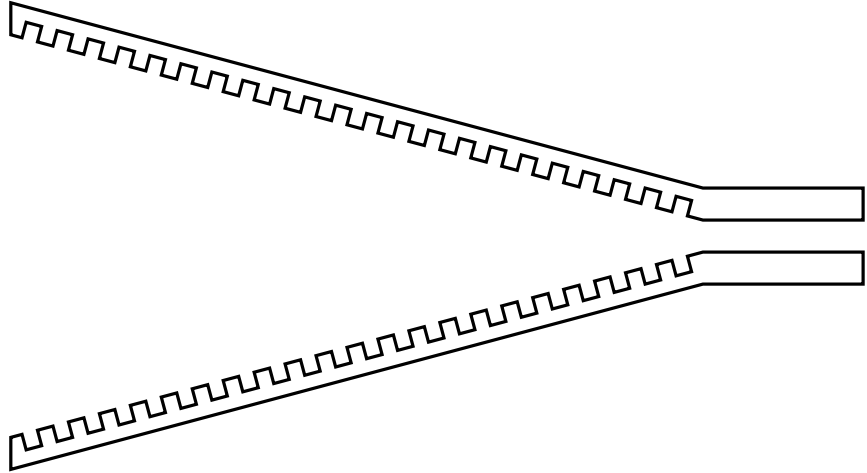}
\caption{Cross-section of corrugated horn.  
The grooves in the walls of the
horn act like shorted $\lambda/4$ stubs, 
and as such appear like open circuits
to the waves propagating in the horn.  
As a result the modes in the horn
are very similar to free-space TEM modes 
which minimizes the discontinuity
at the edge of the horn, and hence minimizes the sidelobes.
\label{fig:corrugated}}
\end{figure}

Shortly after the Cosmic Microwave Background (CMB) was discovered,
the first anisotropy in the CMB was seen:
the dipole pattern due to the motion of the observer relative to
the rest of the Universe (\cite{conklin:1969}).
After confirmation by \cite{henry:1971} and by \cite{corey/wilkinson:1976}
the fourth ``discovery'' of the dipole (\cite{smoot/gorenstein/muller:1977})
showed a very definite cosine pattern as expected for a Doppler effect,
and placed an upper limit on any further variations in $T_{CMB}$.
Further improvements in the measurement of the dipole anisotropy
were made by the Differential Microwave Radiometers (DMR) experiment
on COBE (\cite{bennett/etal:1996} and by the Wilkinson Microwave
Anisotropy Probe (\cite{bennett/etal:2003b}).  
Both the DMR and \iMAP\ use corrugated horns to reduce sidelobes, as shown
in figure \ref{fig:corrugated}.
Everyone of these
experiments used a differential radiometer which measured the difference
between two widely separated spots on the sky.

\begin{figure}[ht]
\plotone{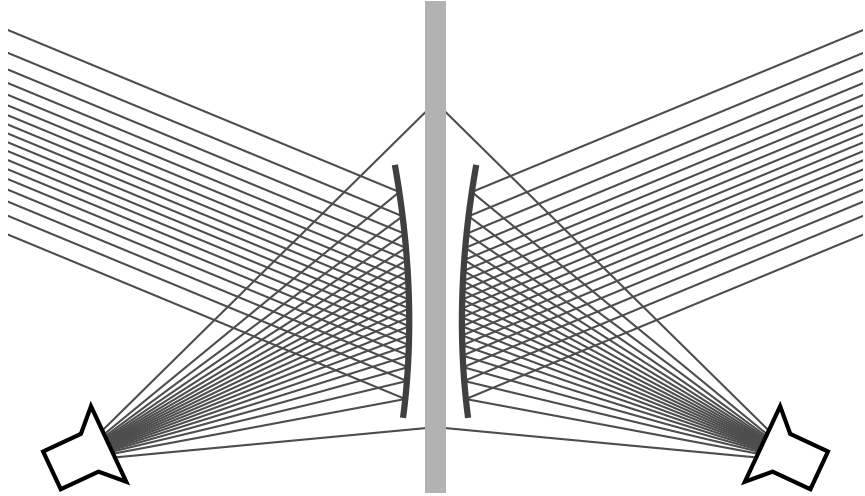}
\caption{A differential radiometer using back-to-back off-axis
paraboloidal dishes.  The feeds are designed to mainly illuminate the
center of the dishes with minimal spillover past the edges.
\label{fig:edge-taper}}
\end{figure}

\begin{figure}[ht]
\plotone{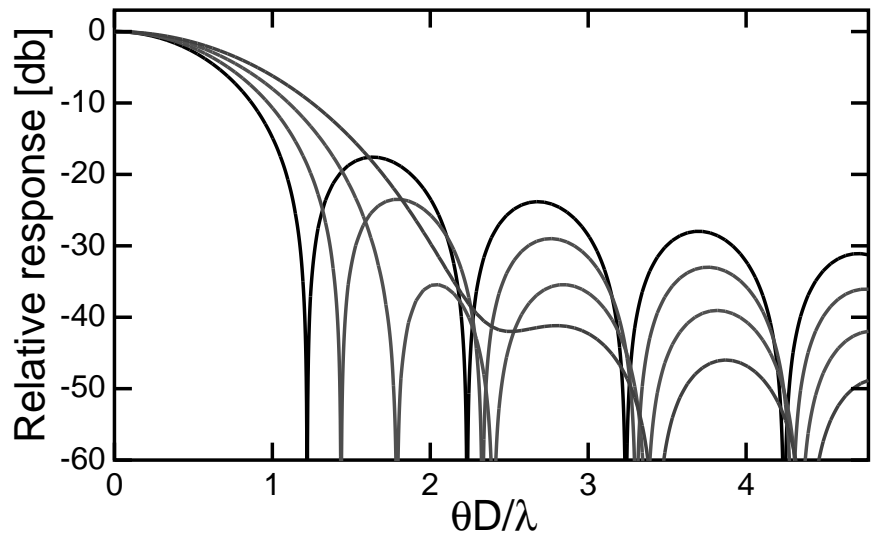}
\caption{The point spread function of a circular aperture for
4 different values of the edge taper with Gaussian illumination.
The four curves are for uniform illumination or 0 db taper, 9, 18
and 27 db taper.  The sidelobe level decreases with increasing taper,
while the width of the main beam increases slightly.
\label{fig:PSF-vs-edge-taper}}
\end{figure}

Experiment to measure smaller angular scales use radio telescopes with
dishes to make a beam with a smaller angular spread than a horn.
Horns would be used to feed the dishes.  A large edge taper should be used
to avoid having the beam from the feed spill over the edge of the dish,
as seen in Figure \ref{fig:edge-taper}.
Usually there is stuff behind the dish that one would rather not look at,
such as the ground or the thermal radiator system in \iMAP.
Figure \ref{fig:edge-taper} illustrates a Gaussian illumination of the
primary with the edge of the dish at the 2$\sigma$ point, which corresponds 
to an edge taper of $e^{-2}$ or about 9 db.
Figure \ref{fig:PSF-vs-edge-taper} shows that sidelobes of a circular
aperture with a Gaussian illumination pattern get much smaller for
increasing edge taper, and that the angular resolution only declines
slightly.
\iMAP\ used edge tapers of 13 to 21 db, although the illumination
patterns were not symmetric or Gaussian (\cite{page/etal:2003}).

The first theoretical predictions of $\Delta T/T = 10^{-2}$
(\cite{sachs/wolfe:1967}) and $\Delta T/T = 10^{-3.5}$
(\cite{silk:1968}) were superseded by predictions based
on cold dark matter (\cite{peebles:1982}, \cite{bond/efstathiou:1987}).
These CDM predictions were consistent with the small anisotropy seen by
COBE and furthermore predicted a large peak at a particular angular scale
due to acoustic oscillations in the baryon/photon fluid prior to
recombination.
The position of this big peak and other peaks in the angular power spectrum
of the CMB anisotropy depends on a combination of the density parameter
$\Omega_m$ and the vacuum energy density $\Omega_V$,
so this peak provides a means to determine the density
of the Universe (\cite{jungman/etal:1996}).
A tentative detection of the big peak at the position predicted
for a flat Universe had been made by 1994 (\cite{scott/silk/white:1995}).
The peak was localized to
$\ell_{pk} = 229 \pm 8.5$ (\cite{knox/page:2000})
by the beginning of 2000.  Later the
BOOMERanG group claimed to have made a dramatic improvement in this
datum to $\ell_{pk} = 197 \pm 6$ (\cite{debernardis/etal:2000}).  This
smaller value for $\ell_{pk}$ favored a moderately closed model for
the Universe.
But improved data on the peak position from \iMAP\ (\cite{page/etal:2003c})
gives $\ell_{pk} = 220.1 \pm 0.8$ which is consistent with a flat
$\Lambda$CDM model.

Polarization of the CMB was shown to be $< 300\;\mu$K
(\cite{lubin/smoot:1981}).
This observation used a differential polarimeter that was only
sensitive to linear polarization.  
COBE put a limit of $< 15\;\mu$K on the polarization anisotropy.
The linear polarization of
the CMB was first detected by DASIPOL (\cite{kovac/etal:2002}),
and the cross-correlation of the temperature and polarization
anisotropies was confirmed by \iMAP\ (\cite{kogut/etal:2003}).

The detected polarization level is an order of magnitude lower than the
anisotropy.  The observed polarization is caused by electron scattering
during the late stages of recombination on small angular scales
and after reionization on large angular scales.
The magnitude of the polarization on small angular scales
depends on the anisotropy being in place
at recombination, as is the case for primordial adiabatic perturbations
but not for topological defects; the electron scattering cross-section; and
the recombination coefficient of hydrogen.  The detection of this polarization
is a very strong confirmation of the standard model for CMB anisotropy.

Because polarization is a vector field, two distinct modes or
patterns can arise (\cite{kamionkowski/kosowsky/stebbins:1997},
\cite{seljak/zaldarriaga:1997}): the gradient of a scalar field (the ``E''
mode) or the curl of a vector field (the ``B'' mode).  Electron scattering
only produces the E mode.  Electron scattering gives a polarization pattern
that is correlated with the temperature anisotropy, so the E modes can be
detected by cross-correlating the polarization with the temperature.
The B modes cannot be detected this way, and the predicted level of the B
modes is at least another order of magnitude below the E modes, or two
orders of magnitude below the temperature anisotropy.

\section{Observational Techniques}

The most important part of any CMB experiment is the modulation scheme that
allows one to measure $\mu$K signals in the presence of $\sim 100\;\mbox{K}$
instrumental foregrounds.  
A good modulation scheme is much more important than high 
sensitivity, since detector noise can always be beaten down as $1/\sqrt{t}$
by integrating longer, while a systematic error is wrong forever.

\subsection{Chopping}

\begin{figure}[ht]
\plotone{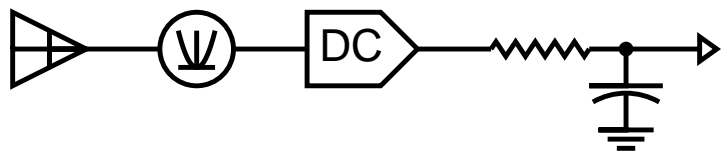}
\caption{A total power radiometer with a bolometer [a square-law device
indicated by the parabolic curve] feeding a DC
amplifier.  This design will have too much $1/f$ noise to be useful.
\label{fig:DC-bolometer}}
\end{figure}

The first step in any modulation scheme is the chopping scheme.  The
instrument sketched out in Figure \ref{fig:DC-bolometer} will not
succeed because the first stage of amplification is at 
zero frequency (DC),
and all electronic circuits suffer from either $1/f$ noise or
drifting baselines corresponding to $1/f^2$ noise.
Figure \ref{fig:DC-bolometer} shows a bolometer detector where 
the radiation goes directly into a square-law device.  In terms
of radio engineering, this is similar to the crystal sets that
were used in the 1910's.  Modern bolometers running at temperatures
below $0.3\;\mbox{K}$ actually have enough sensitivity to make this
design superior to radio frequency amplifier designs, but
some form of chopping is absolutely required.

\begin{figure}[ht]
\plotone{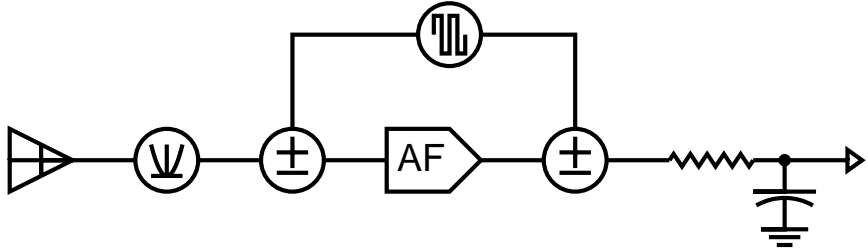}
\caption{A total power radiometer with an AC-biased bolometer feeding
an audio amplifier followed by a lockin amplifier.
This design is used in
BOOMERanG and will be used in the Planck HFI.
\label{fig:AC-bolometer}}
\end{figure}

The least obtrusive chopping scheme involves biasing the bolometer with
alternating current (AC).  This is illustrated in Figure
\ref{fig:AC-bolometer}.  The bias supply is connected to an audio
frequency (AF) source, shown here as a square wave oscillator, and this
causes the responsivity of the bolometer to change sign at an audio
frequency rate.  The output of the bolometer then goes through an AF
amplifier and into a phase sensitive demodulator and low pass filter,
or a lockin amplifier.  The $1/f$ knee of an AC-biased bolometer can be
lower than 0.01 Hz (\cite{wilbanks/etal:1990}).  While AC bias removes
the problem of $1/f$ noise due to the amplifier, there can still be
$1/f$ or $1/f^2$ noise from the atmosphere or drifting temperatures in
the instrument.  Thus a good scanning strategy is still needed with
AC-biased bolometers.

\begin{figure}[ht]
\plotone{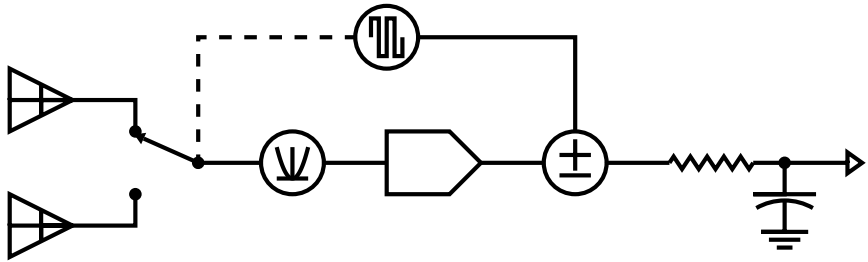}
\caption{A differential radiometer with an optical chopper
so the bolometer looks alternately at two sky different sky
positions.  The bolometer feeds
an audio amplifier followed by a lockin amplifier.
\label{fig:SPDT-bolometer}}
\end{figure}

A differential radiometer like the COBE DMR looks alternately at two
different sky positions and measures the difference between the
brightnesses at these two positions.  Figure \ref{fig:SPDT-bolometer}
shows a differential radiometer with a bolometric detector.
This system using a chopping secondary is fairly common in infrared
astronomy.

\begin{figure}[ht]
\plotone{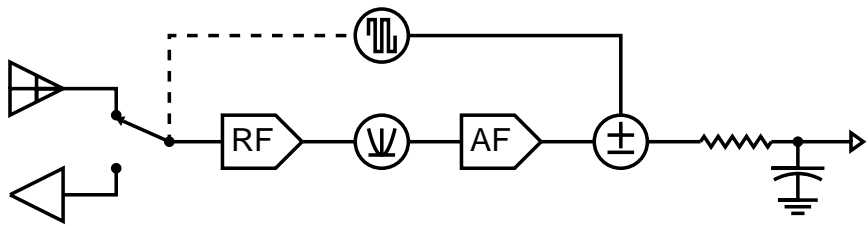}
\caption{A differential radiometer chopping against a load, using
an RF amplifier prior to the square-law detector.
This kind of instrument is required when measuring the absolute
temperature of the CMB, and was used by \protect\cite{penzias/wilson:1965}.
\label{fig:SPST-HEMT}}
\end{figure}

Figure \ref{fig:SPST-HEMT} shows a radiometer using a radio frequency
(RF) amplifier that is chopping against a load.  One might think that with
the first stage of amplification occuring at a high frequency, chopping
would not be necessary, but in practice RF amplifiers have gain
fluctuations that contribute multiplicative $1/f$ noise.  Chopping
against a load is necessary when measuring the absolute temperature
of the CMB, $T_\circ$.

In terms of antique radio technology, this radiometer with an RF amplifier
leading to a square-law device is a {\em tuned RF} receiver which was
the state-of-the art in 1929.  The modern superheterodyne circuit for
radio receivers with ampflication and filtering at an intermediate
frequency (IF) was used by the COBE DMR, but the primary advantage of a
superheterodyne receiver over a tuned RF receiver is its improved
selectivity.  Since the CMB is a very broad band signal, selectivity
beyond that provided by a RF filter is seldom desired.

\begin{figure}[ht]
\plotone{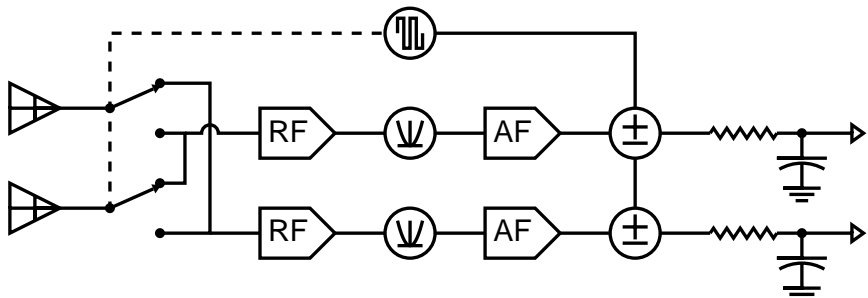}
\caption{A dual channel differential radiometer where the chopper
acts like a reversing switch.  Practical systems with arrays like
SCUBA act like this when running with chopper throws smaller than
the array.\label{fig:DPDT-HEMT}}
\end{figure}

Finally one can set up a chopping system with two antennae and two
amplifier chains, so that the chopper reverses the connections
between the horns and the amplifiers.  Figure \ref{fig:DPDT-HEMT}
shows such a scheme.  In reality this setup would not be very practical,
but the same effect is obtained when using an array of detectors
like SCUBA behind a chopper with a throw that is less than the size
of the array.

\begin{figure}[ht]
\plotone{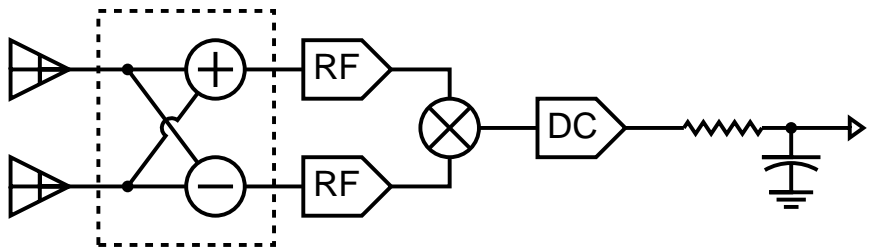}
\caption{A correlation radiometer where the chopper
is replaced by a hybrid circuit. Separate amplifier chains amplify
the sum and difference signals which are then multiplied.
\label{fig:DC-correlation}}
\end{figure}

A practical microwave radiometer that has the same sensitivity as the
system shown in Figure \ref{fig:DPDT-HEMT} is the correlation radiometer
shown in Figure \ref{fig:DC-correlation}.  A hybrid circuit at the input
forms the sum and difference voltages $(V_A+V_B)/\sqrt{2}$ and
$(V_A-V_B)/\sqrt{2}$.  
These are separately amplified and then multiplied,
giving and output proportional to $V_A^2-V_B^2$ which is the desired
difference in the powers arriving at the two horns.

\begin{figure}[ht]
\plotone{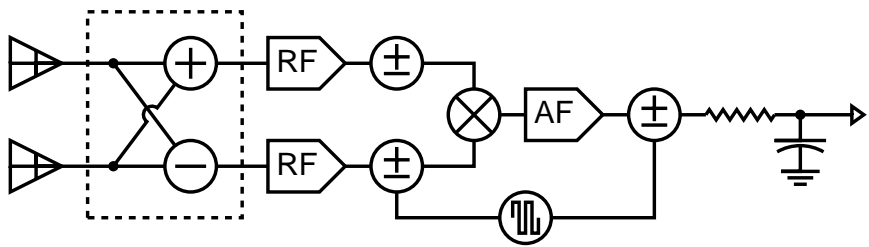}
\caption{A correlation radiometer with a phase switch inserted
in both arms.  One of the phase switchs is driven to modulate
the output of the multiplier so a lockin amplifier can be used.
\label{fig:AC-correlation}}
\end{figure}

There are two practical difficulties with the correlation radiometer.
The first is that one will get too much $1/f$ noise from the multiplier
and the DC amplifier following it.  This can be solved by introducing
phase switches into both amplifier chains, and then toggling just one
of them.  This causes the sign of the output product to toggle,
so the output of the multiplier can be amplified at audio frequencies
and then fed to a lockin amplifier.  This is shown in
Figure \ref{fig:AC-correlation}.

The second practical difficulty is the implementation of the multiplier
needed for the correlation radiometer.  This multiplier has to form
a product in a picosecond in order to handle the $94\;\mbox{GHz}$
output of the highest frequency \iMAP\ band.  This is handled by
using another hybrid followed by square law detectors, which corresponds
to using a {\em quarter-square multiplier}.  The ancient Egyptians used
tables of squares to multiply using the formula 
$A\times B = [(A+B)^2 - (A-B)^2]/4$.  Both \iMAP\ and the Planck LFI
use this technique, giving what is termed a differential pseudo-correlation
radiometer.

These chopping schemes have an effect on the signal-to-noise ratio of
the experiment as follows:
\begin{itemize}
\item Let the mapping speed of the total power, one-horned system
like Figure \ref{fig:AC-bolometer} be 1.00.
This is the system planned for the Planck HFI.
\item Then the system chopping against a load like 
Figure \ref{fig:SPST-HEMT}
spends only 50\% of its time looking at the sky, so the sky signal is
$\sqrt{2}$ noisier.  The noise on the load measurement is also
$\sqrt{2}$ noisier because the load is observed only 50\% of the time.
The difference output is then 2 times noisier, which corresponds to
a mapping speed of only 0.25 relative to the total power system.
\item The two-horned differential radiometer like
Figure \ref{fig:SPDT-bolometer} is looking at
two parts of the sky at once,
so it has a mapping speed of 0.5 relative to the total power system.
\item The two-horned differential radiometers with two amplifier chains
like Figure \ref{fig:DPDT-HEMT} or Figure \ref{fig:AC-correlation} achieve
a mapping speed equal to the total power system, at the expense of
doubling the number of horns and amplifiers.  This is the system
used by \iMAP.
\item The Planck LFI is like \iMAP\ but one of the horns is replaced
by a load, so its mapping speed is 0.5 relative to the total power system.
\end{itemize}

\subsection{Scanning}

Any experiment to map $N_p$ pixels will need to collect $N_d \geq N_p$
data points.  One would like to see that a typical time history that
might be produced by some systematic effect will correspond to an
element in the $N_d$-dimensional data space that is orthogonal or
nearly orthogonal to the $N_p$-dimensional subspace that corresponds to
the time histories that can be generated by scanning a map.  This can
be achieved by imposing more than two distinct modulations in the
experiment, since the sky is a two dimensional object.  For example,
the COBE DMR chopped between two beams 100 times per second, spun to
interchange those beams every 73 seconds, precessed that spin axis
around the circle $94^\circ$ away from the Sun every 104 minutes, and
then moved that circle around the sky once per year as the Earth went
around the Sun.  This is a four way modulation.  \iMAP\ chops between
two beams 2500 times per second, spins to interchange those beams every
132 seconds, precesses its spin axis around a circle $157.5^\circ$ from
the Sun once per hour, and follows the annual motion of the Sun again
giving a four way modulation.

On the other hand ARCHEOPS only scanned around a circle of constant
elevation and then let the center of the circle move in right ascension
as the Earth turned.  This provides only a two way modulation.  Since
the sky itself is a two dimensional function, just about any time
history of drifting baselines is consistent with some pattern on the
sky.  Thus ARCHEOPS is very vulnerable to striping.  This can be seen
in the last panel of Figure 2 of astro-ph/0310788
(\cite{hamilton/benoit/etal:2003}) which clearly shows correlated
residuals aligned with the scan path.  These stripes have a low enough
amplitude to not interfere with measurements of the
temperature-temperature angular power spectrum $C_\ell^{TT}$, but they
would ruin a measurement of the polarization power spectrum
$C_\ell^{EE}$.

\begin{figure}[ht]
\plotone{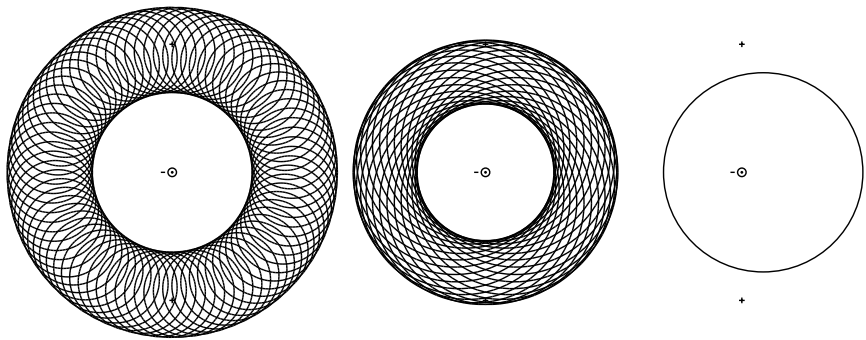}
\caption{The ``hourly'' scan pattern of COBE, \iMAP\ and Planck from
left to right.  In each panel the anti-solar direction is in the
center, and plus signs denote the North and South ecliptic poles.  An
equiangular azimuthal projection is used.  COBE scanned this area in
one orbit of about 103 minutes, \iMAP\ scans its pattern every hour,
while Planck spends several hours integrating on one scan circle of
radius $70^\circ$ radius shown offset from anti-Sun by a $15^\circ$
precession angle.  \label{fig:scan-patterns}}
\end{figure}

Stripes are caused by small, asymmetric reference sets for pixels in
the map.  The reference set for the $i^{th}$ pixel consists of the
other pixels in the map that are used to establish the baseline for the
$i^{th}$ pixel.  In a differential experiment like COBE or \iMAP\ the
reference set is the circle of radius equal to the chopper throw
centered on the $i^{th}$ pixel, or a subset of this circle.  This gives
a large reference set so differential experiments have nearly
uncorrelated noise per pixel and thus no stripes.

\begin{figure}[ht]
\plotone{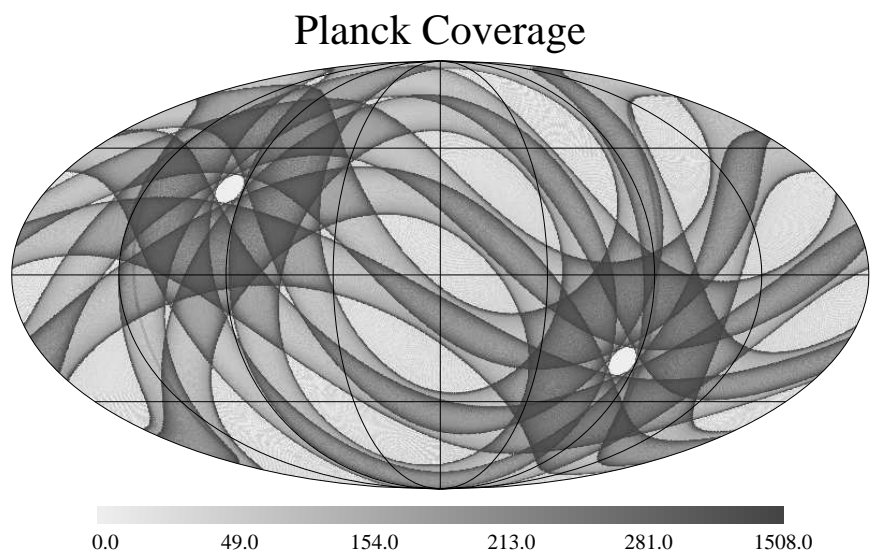}
\caption{Simulated coverage by the Planck mission assuming a $70^\circ$
scan radius and a $15^\circ$ precession radius with 10 precession
cycles per year.  This plot is in galactic coordinates.  Note the
asymmetry between the coverage patterns in the North and south ecliptic
hemispheres.  \label{fig:Planck-coverage}}
\end{figure}

For a one-horned experiment like ARCHEOPS or Planck the reference set
is a line of pixels passing through the $i^{th}$ pixel along the scan
direction.  The length of the reference set along the scan circle 
is determined by the $1/f$ knee of the output, and is of order
$\pm\omega/f_{knee}$ where $\omega$ is the angular scan rate of the
instrument.  
Observations both before and after the $i^{th}$ pixel
can be used to set the baseline so the reference set always has
inversion symmetry.
A description of the minimum variance method for processing
data from one-horned radiometers using a ``pre-whitening'' filter
and time-ordered processing techniques is given by
\cite{wright:1996}.
The width of the pre-whitening filter determines the length of
the reference set.
When several scan circles pass through the $i^{th}$ pixel
in different directions then the reference set becomes larger and more
symmetric.  If scans pass through the $i^{th}$ pixel in all directions
(modulo $180^\circ$ because of the inversion symmetry) then the reference
set is symmetric and there are no stripes.
If $\omega/f_{knee}$ is large and there is a large range of scan angles
then the reference set is large and
the noise per pixel is nearly uncorrelated.

\begin{figure}[ht]
\plotone{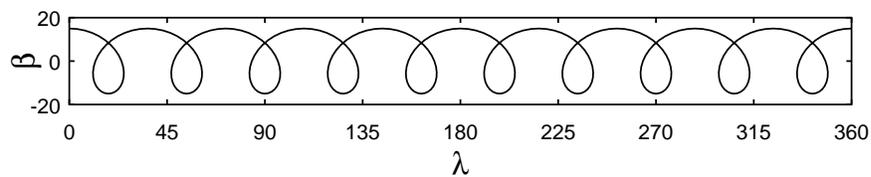}
\caption{The path of the Planck scan circle center when the precession
angle is $15^\circ$ and the precession rate is 10 cycles per year.
The difference between the Northern and Southern loops of the cycloid
causes the asymmetry in the coverage pattern seen in Figure
\protect\ref{fig:Planck-coverage}.
\label{fig:Planck-cycloid}}
\end{figure}

Figure \ref{fig:scan-patterns} shows the amount of sky scanned by COBE, 
\iMAP\ and Planck in about an hour.  These scan patterns move around the
sky once per year as the Earth orbits the Sun.  The center of the
Planck scan circle precesses around the anti-Sun slowly, with perhaps
10 turns per year.  The range of scan angles through a pixel
is always $180^\circ$ for the COBE scan pattern.  For the \iMAP\ pattern
the region near the ecliptic only sees a range of $45^\circ$ in scan
angles, while the median range of scan angles is 
$95^\circ$.  The median range of scan angles is only $51^\circ$ for Planck.

The range of scan angles through a pixel is also crucial in determining
the ability of a given experiment to make reliable measurements of
polarization.  Observing the same pixel with different orientations of
the instrumental axes provides the data needed to separate true celestial
polarizations from instrumental effects.

The simulated Planck coverage map created while determining the range of
scan angles, shown in Figure \ref{fig:Planck-coverage}, illustrates an
interesting asymmetry that is inherent in this mission's
planned slow precession.  If Planck precesses only 10 times per year
on a $15^\circ$ radius circle, then the scan circle motion due to
precession is only 2.5 times higher than the rate due to the annual
motion.  Thus in one ecliptic hemisphere the net motion is 3.5 times
the annual rate while in the opposite hemisphere the net motion is
only 1.5 times the annual rate.
This results in an asymmetric path for the scan circle center shown
in Figure \ref{fig:Planck-cycloid}.
This asymmetry will certainly make testing any North-South asymmetry
(\cite{eriksen/etal:2003})
much more difficult.
Planck would have a much better scanning strategy if the precession rate
were close to the geometric mean of the spin period and one year.  This
would be about 10 hours, or several hundred precession periods per year.

\subsection{Frequency Range}

One time history that will always be consistent with a pattern on the
sky is obtained by scanning over the Milky Way.  The only way to make
this nearly orthogonal to a true CMB pattern on the sky is to observe
a large range of frequencies.  The spectrum of the Milky Way on large
angular scales as measured by FIRAS is given in \cite{wright/etal:1991}.
The ratio between the CMB anisotropy signal and this galactic spectrum
peaks at $72\;\mbox{GHz}$.  For higher frequencies the rising thermal
dust emission spectrum starts to dominate over the CMB signal.
At frequencies lower than $72\;\mbox{GHz}$ the galactic foreground
is dominated by free-free and synchrotron emission.
An experiment to measure the primary CMB anisotropy would like to observe
a range of frequencies covering $\pm$ a factor of three from $72\;\mbox{GHz}$,
or from $24\;\mbox{GHz}$ to $216\;\mbox{GHz}$
But the thermal Sunyaev-Zeldovich effect goes through zero at about
$220\;\mbox{GHz}$ so extending the high frequency limit to $400\;\mbox{GHz}$
is clearly a good idea.
The \iMAP\ mission only covers the peak and the low frequency side
of the peak in the CMB:galaxy ratio, while the Planck mission will extend
the high frequency coverage to more than $800\;\mbox{GHz}$.

\subsection{Sensitivity}

Once a good chopping and scanning strategy is planned, a detector system
with enough sensitivity to map the CMB anisotropy is needed.  The
primary anisotropy of the CMB extends up to $\ell \approx 2000$ so there
are about 4 million spots on the sky that need to be measured.  The
anisotropy is about $38\;\mbox{$\mu$K}$ in each spot so the integrated
``monopole'' sensitivity, $\sigma_{pix}/\sqrt{N_{pix}}$, needs to be
about $19\;\mbox{nK}$ in order to reach a signal-to-noise ratio of
1 per spot on the primary anisotropy.  A SNR of 1 marks the
``point of diminishing returns'' when measuring the variance of
a Gaussian signal.  
When the SNR per pixel is $< 1$ then the error on $C_\ell$ improves
like one over the integration time, while when the SNR per pixel
is $> 1$ then the error on $C_\ell$ is limited by cosmic variance
and does not improve at all with increased integration time.
But to measure E-mode polarization one would like 10 times
more sensitivity, and to measure the B-mode polarization one would
like at least 100 times more sensitivity.

\iMAP\ will achieve a monopole sensitivity of $23\;\mbox{nK}$
in 4 years which is well into the region of diminishing returns
on $C_\ell^{TT}$ for the $\ell < 900$ range compatible with the
\iMAP\ angular resolution.  But the \iMAP\ sensitivity is far from
the point of diminishing returns for polarization measurements.

How can one reach these sensitivity goals?  The goal of a monopole
sensitivity of $19\;\mbox{nK}$ can be achieved in one year with a
sensitivity of $107\;\mbox{$\mu$K}$ in one second.  With a bandwidth
of $18\;\mbox{GHz}$ (25\% of the $72\;\mbox{GHz}$ optimal frequency)
the system temperature requirement is $T_{sys} = 14\;\mbox{K}$
for a single radiometer channel, using the Dicke radiometer
equation $\Delta T = T_{sys}/\sqrt{Bt}$.  The best current
performance of High Electron Mobility Transistor (HEMT)
amplifiers is about $0.3\;\mbox{K/GHz}$ for cryogenic HEMTs,
which is a bit too high.  Hence an experiment designed to map the
whole sky to the point of diminishing returns for $C_\ell^{TT}$
would need to have at least two channels.
\iMAP\ has 20 channels with two polarizations on each of the 10 
differencing assemblies, but only achieves $1.5\;\mbox{K/GHz}$
with passively cooled HEMTs at running at $\approx 90\;\mbox{K}$.
However, the absence of expendable cryogens allows \iMAP\ to
operate for several years and easily surpass the sensitivity goal.

The $1/f$ gain fluctuations in HEMTs require a high chopping frequency.
Prior to the lockin amplifier, the variance of the output 
of a HEMT radiometer in a $1\;\mbox{Hz}$ bandwidth 
($t_{int} = 0.5\;\mbox{sec}$) centered at $f$ is given by
\begin{equation}
\mbox{var}(\Delta T) = \frac{2 T_{sys}^2}{B}
+\left(\frac{\Delta G}{G}(f)\right)^2 T_{sys}^2
\end{equation}
The $1/f$ gain fluctuations are given by
\begin{equation}
\left(\frac{\Delta G}{G}(f)\right)^2 = 
b^2 \left(\frac{1 \;\mbox{Hz}}{f}\right)^\alpha
\end{equation}
Typically $\alpha \approx 1$ and
$b = 10^{-5}/\sqrt{\mbox{Hz}}$ for warm HEMTs 
and $10^{-4}/\sqrt{\mbox{Hz}}$ for cryogenic
HEMTS, and the bandwidth $B$ is 10's of GHz, so the $1/f$
knee frequency is 
\begin{equation}
f_{knee} = (b^2 B/2)^{1/\alpha}
\end{equation}
which ranges from 20 to 1000 Hz for the \iMAP\ radiometers
(\cite{jarosik/etal:2003}).
The chopping frequency $f_c$ must be higher than this to 
avoid excess noise due to gain fluctuations.

The post-lockin noise variance in 1 Hz centered at $f$ is
\begin{equation}
\mbox{var}(\Delta T) = 4 \, T_{sys}^2 \left[\frac{2}{B}
+\left(\frac{\Delta G}{G}(f_c)\right)^2\right]
+T_{off}^2 \left(\frac{\Delta G}{G}(f)\right)^2
+\Delta T_{off}(f)^2
\end{equation}
which still shows $1/f$ noise due to gain
fluctuations but they are only driven by the imbalance in the
radiometer, $T_{off}$.  The factor of ``4'' in front of
$T_{sys}^2$ is the increased noise due to chopping.
If $T_{off} << T_{sys}$ then the knee frequency is much lower:
\begin{equation}
f_{knee}^\prime = \left(\frac{b^2 \, B \, T_{off}^2}
{8 \, T_{sys}^2 \, [1+(f_{knee}/f_c)^\alpha]}
\right)^{1/\alpha}.
\end{equation}
This is 0.04 Hz for the worst case \iMAP\ radiometer, W4,
which has both the highest bandwidth and the highest
offset.  Ideally the post-lockin $1/f$ knee frequency
$f_{knee}^\prime$ should be lower than all the scan
frequencies but the \iMAP\ spin frequency is only
0.008 Hz so this ideal was not achieved for the W4.
$\Delta T_{off}(f)^2$ is the power spectrum of the offset 
drifts which typically show $1/f^2$ behavior and dominate
the noise at very low frequencies.

Note that the quantum limit on coherent receivers is $0.5 h\nu/k$
or $0.024\;\mbox{K/GHz}$ (\cite{wright:1999}).  This corresponds to
0.5 photons per mode.  The $0.3\;\mbox{K/GHz}$ for cryogenic HEMTs
is about $7\;\mbox{photons/mode}$.  At $72\;\mbox{GHz}$ the CMB has
only $\overline{n} = 0.4$ photons per mode,
where $\overline{n} = (\exp[h\nu/kT]-1)^{-1}$ is the mean number
of photons per mode.
Thus a background-limited incoherent detector could be much more
sensitive than a coherent radiometer using HEMTs.

\begin{figure}[ht]
\plotone{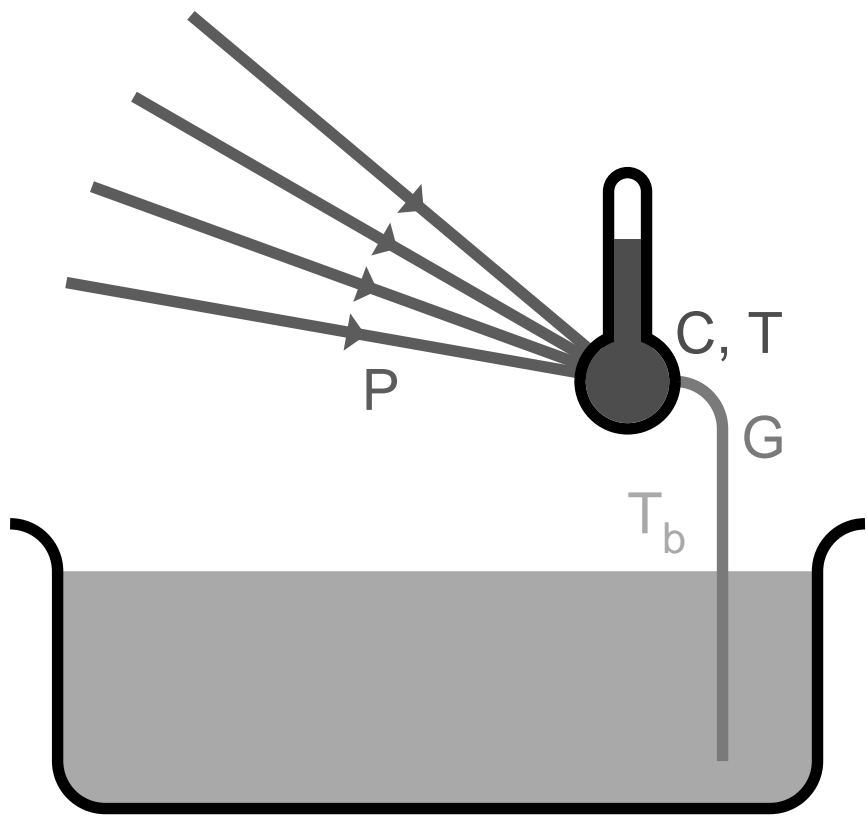}
\caption{A cartoon representation of a bolometer: a small
thermometer at the focus of a beam of radiation, thermally
linked to a bath.
\label{fig:bolometer-cartoon}}
\end{figure}

Consider a bolometric radiometer with an $18\;\mbox{GHz}$ at
$72\;\mbox{GHz}$ with a diffraction-limited throughput, $A\Omega =
\lambda^2$.  A background limited (BLIP) system would have a
temperature sensitivity of $19\;\mbox{$\mu$K}$ in 1 second.  This is
already 6 times better than the $107\;\mbox{$\mu$K}$ in one second
needed to reach the point of diminishing returns for $C_\ell^{TT}$ with
a single channel.  However, this bolometer would have to have a noise
equivalent power (NEP) less than $7 \times
10^{-18}\;\mbox{W/$\sqrt{Hz}$}$, which is still difficult to achieve.
The required NEPs are higher and thus easier to achieve at higher
frequencies so bolometers are definitely the technology of choice for
frequencies higher than $94\;\mbox{GHz}$.

\begin{figure}[ht]
\plotone{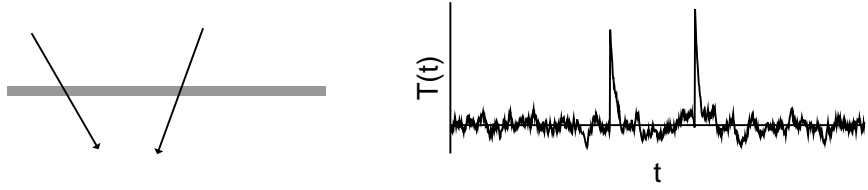}
\caption{The bolometer shown at left responds to the energy deposited
by ionizing particles passing through its absorber, leading the
the impulsive signals known as glitches shown at right.
\label{fig:bolometer-CRs-and-glitches}}
\end{figure}

\begin{figure}[ht]
\plottwo{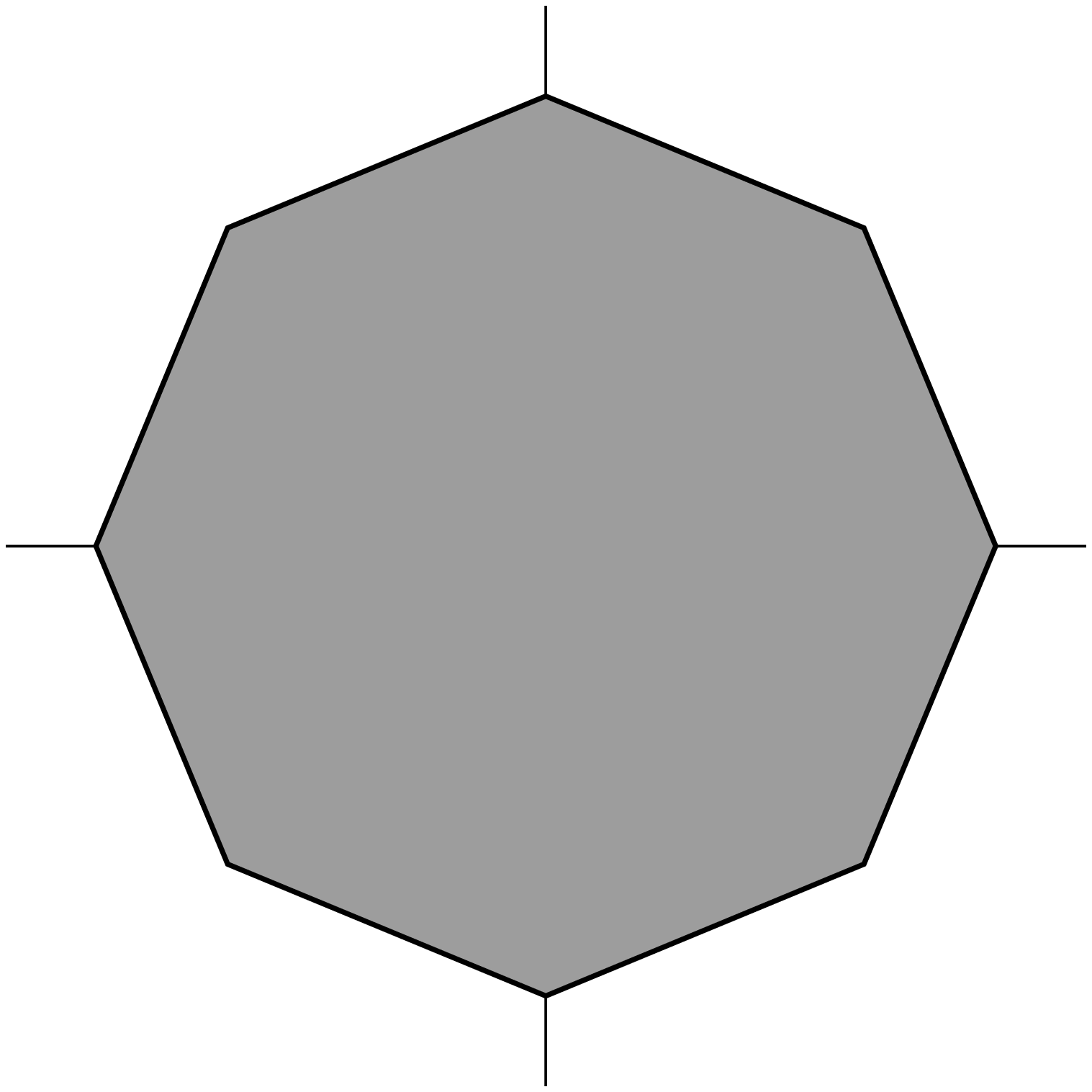}{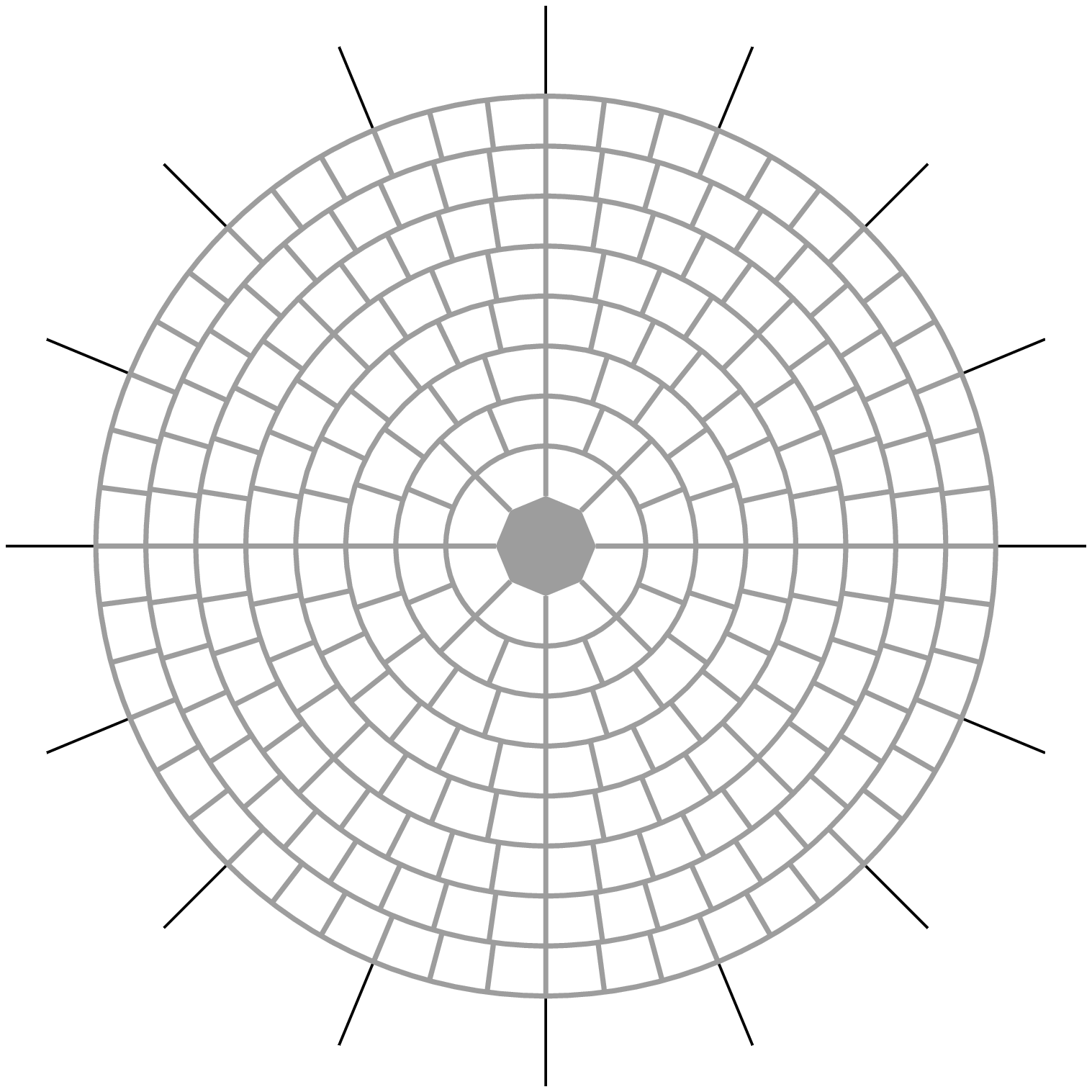}
\caption{The FIRAS bolometer on the left compared to a spiderweb
bolometer on the right.
\label{fig:FIRAS-spider}}
\end{figure}

Bolometers are just thermometers weakly coupled to a thermal bath
by a conductance $G$.  
Radiation is focussed on the bolometer
causing its temperature to rise by $\Delta T = P/G$ 
(see Figure \ref{fig:bolometer-cartoon}).
The thermal
time constant of a bolometer is given by $\tau = C/G$ where 
$C = dQ/dT$ is
the heat capacity of the bolometer in ergs/K.  From the definition
of entropy $S = k \ln\Omega$ with $\Omega$ being the state density,
and $dS = dQ/T$, we find that 
\begin{equation}
\ln\Omega_{bolo} = \ln\Omega_\circ + (Q-Q_\circ)/kT_b
-0.5(Q-Q_\circ)^2/(kT_b^2 C) + \ldots
\end{equation}
where $Q_\circ$ is the energy of the bolometer at the bath
temperature $T_b$, which gives
\begin{equation}
\frac{d\ln\Omega}{dQ} = \frac{1}{kT} = \frac{1}{kT_b}\left(1 - 
\frac{Q-Q_\circ}{CT_b}\right) + \ldots
\end{equation}
and thus $T = T_b + (Q-Q_\circ)/C + \ldots$ as required.
The thermal bath has a much larger heat capacity so the
overal density of states is
\begin{equation}
\Omega_{bolo}\Omega_{bath} \propto \exp[-0.5(Q-Q_\circ)^2/(kT_b^2C)]
\end{equation}
which is a Gaussian with a standard deviation of the energy in the
bolometer of $\sigma(Q) = T\sqrt{kC}$.
This corresponds to a standard deviation of the power
$\sigma(P) = \sigma(Q)/\tau$ and since the noise bandwidth of
a simple lowpass filter with time constant $\tau$ is $1/4\tau$, a
noise equivalent power of
$\mbox{NEP} = T\sqrt{4kC/\tau}$.

Clearly one obtains the best performance for a given time constant with
a detector that has the lowest possible heat capacity.  The heat
capacity of a crystal varies like $C \propto (T/\Theta_D)^3$, where
$\Theta_D$ is the Debye temperature.  Diamond has the highest Debye
temperature of any crystal, so FIRAS used an 8 mm diameter, 25 $\mu$m
thick disk of diamond as a bolometer
(\cite{mather/fixsen/shafer:1993}).  Diamond is transparent, so a very
thin layer of gold was applied to give a surface resistance close to
the 377 ohms/square impedance of free space.  On the back side of the
diamond layer an impedance of 267 ohms/square gives a broadband
absorbtion.  Chromium was alloyed with the gold to stabilize the
layer.  The temperature of the bolometer was measured with a small
silicon resistance thermometer.  Running at $T = 1.6$~K, the FIRAS
bolometers achieved an optical NEP of about $10^{-14}$~W/$\sqrt{Hz}$.

Since $C \propto T^3$ and $\mbox{NEP} = T\sqrt{4kC/\tau}$
the NEP scales like $T^{2.5}$.  This means that a FIRAS-like
bolometer running at 0.1 K could achieve an NEP of 
$10^{-17}$~W/$\sqrt{Hz}$.

A bolometer is sensitive to any source of heat, not just
microwave photons, so charged particles passing through the
absorbing layer lead to impulsive signals called glitches
as seen in Figure \ref{fig:bolometer-CRs-and-glitches}.
These events occur most frequently at the stratospheric altitudes
where balloon-borne experiments operate.  An important
improvement in bolometer design was the use of mesh absorbers,
since there is no need to fill an area $\sim \lambda^2$ in order
to absorb radiation with wavelength $\lambda$.  
Figure \ref{fig:FIRAS-spider} shows how the area that is sensitive 
to charged particles can be cut using a spiderweb bolometer
(\cite{bock/etal:1995}).  This cuts the mass and hence the
heat capacity of the absorber.  BOOMERanG used spiderweb bolometers.

Antenna-coupled bolometers (\cite{schwarz/ulrich:1977}) offer another
way to achieve a small heat capacity and a small area sensitive to
charged particles.  Radiation is absorbed by an antenna and then
coupled into a transmission line which brings it to a very small
absorbing thermometer.

\section{Recent Observations}

In this paper I will discuss the new observations that have been released in
the year prior to this meeting: September 2002 to September 2003.
I will discuss these results in time order.

\subsection{DASIPOL}

The Degree Angular Scale Interferometer (DASI) is a very small 
interferometric array that
operates at 26-36 GHz and the South Pole.  After measuring the angular
power spectrum of the anisotropy (\cite{halverson/etal:2002})
the instrument was converted into
a polarization sensitive interferometer which detected the E mode polarization
at 5.5$\sigma$ by looking at a small patch of sky for most of a year of
integration time (\cite{kovac/etal:2002}).
The level agreed well with the solid predictions for adiabatic primordial
perturbations.  Since the measured quantity was the EE autocorrelation, the
5.5$\sigma$ corresponds to a 9\% accuracy in the polarization amplitude.

The TE cross-correlation was also seen, but with only 50\% accuracy.
As expected, the B modes were not seen.

\subsection{ARCHEOPS}

ARCHEOPS is a balloon-borne experiment built to test the detectors and the
cryogenic system planned for the ESA Planck Explorer mission High Frequency
Instrument. ARCHEOPS has bolometers cooled below 0.1 K, and thus achieves
a very high instantaneous sensitivity and was able to map a substantial
fraction of the sky with good SNR in only 1 night of observing.
This large sky coverage provided a lower level of cosmic or sample
variance, so the ARCHEOPS data gave much better results on the low-$\ell$
side of $\ell_{pk}$.  This gave a peak location of
$\ell_{pk} = 220 \pm 6$ (\cite{benoit/etal:2003}).

\subsection{ACBAR}

ACBAR is a bolometric camera array mounted on the VIPER
2 meter diameter telescope at the South Pole.  It operates at several
frequencies on both sides of the peak of the spectrum of the CMB, and
thus can be used to make a sensitive test for the Sunyaev-Zeldovich effect.
However, the data released to date are only at one frequency.  
ACBAR was able to measure the CMB angular power spectrum
up to $\ell > 2000$ (\cite{kuo/etal:2002}),
but the size of the surveyed regions was so small that
ACBAR's $\ell$ resolution was too limited to provide much information about
$\ell_{pk}$.  ACBAR was able with its high observing frequency
to show that the excess at $\ell > 2000$ seen by CBI
(\cite{mason/etal:2003}) is not a primary
CMB anisotropy.  It could be due to point sources, or it could be due to
the S-Z effect, both of which would be much weaker in the ACBAR band than
in the CBI 26-36 GHz band.

\subsection{\iMAP}

\begin{figure}[htb]
\plotone{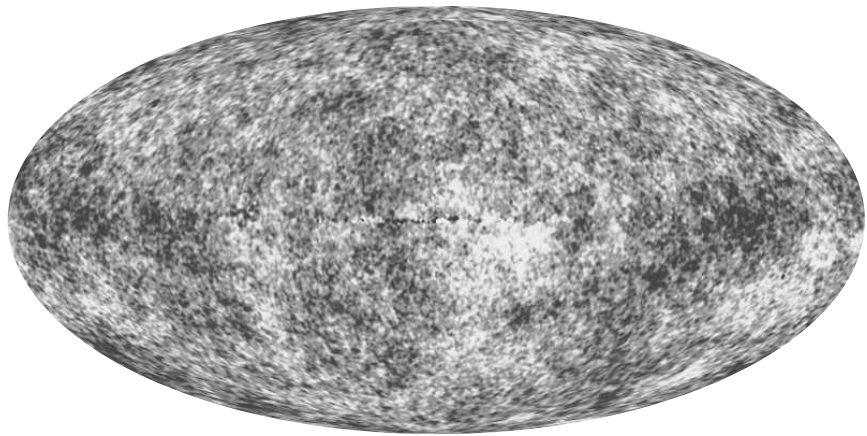}
\caption{A ``no galaxy'' map made from an internal linear combination
of the 5 \iMAP\ bands, smoothed to $1^\circ$ resolution.
\label{fig:ilc}}
\end{figure}

\begin{figure}[htb]
\plotone{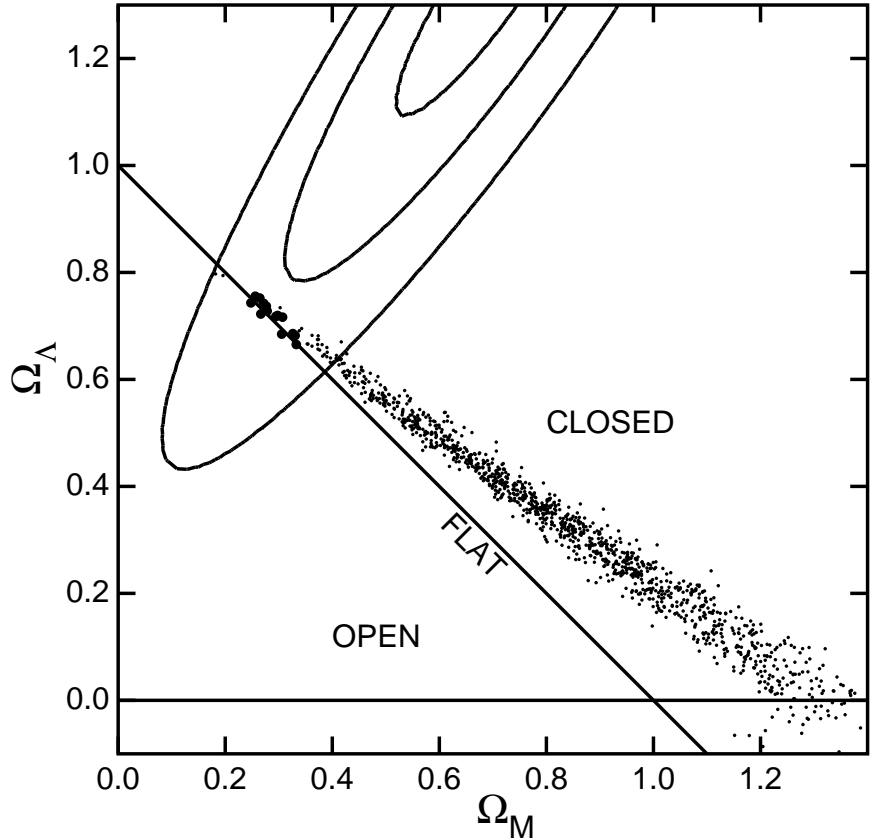}
\caption{Cloud of points from a Monte Carlo Markov chain sampling of the
likelihood of models fit to the \protect\iMAP\ plus other CMB datasets.
The size of the points indicates how consistent the model is with the 
HST Key Project on the Distance Scale value for the Hubble constant.
The contours show the likelihood computed for 230 Type Ia supernovae
(\cite{tonry/etal:2003}).\label{fig:Wv-Wm-wSNe-wMAP}}
\end{figure}

The \iMAP\ satellite, launched on 30 June 2001, released it first year
results on 11 Feb 2003.  Simultaneously the mission was renamed the
Wilkinson Microwave Anisotropy Probe to honor the late David T.
Wilkinson who was a key member of both the COBE and the \iMAP\ teams
until his death in September 2002.  

\iMAP\ observed at 5 frequencies: 23, 33, 41, 61 and 94 GHz.  From 
maps in these 5 bands, an internal linear combination map has been 
constructed that cancels almost all of the Milky Way foreground while
preserving the CMB anisotropy.  Figure \ref{fig:ilc} shows this map on
a gray scale.  All bands were smoothed to $\approx 1^\circ$ resolution so the
linear combination could be made without worrying about the different
beamsizes in the different bands.  After this smoothing 
53\% of the
sky was within $\pm 53\;\mu$K of the median of the map, implying an RMS
$\Delta T$ of 73 $\mu$K in this smoothing beam.  
This is considerably higher than
the 30 $\mu$K RMS seen by the COBE DMR at a $10^\circ$ smoothing because
a $1^\circ$ beam picks up a large part of the big first acoustic peak.

The \iMAP\ results and their
cosmological significance were described in 13 papers and will not be
repeated here.  \cite{bennett/etal:2003} gave a description of the
\iMAP\ mission.  \cite{bennett/etal:2003b} summarized the results from
first year of \iMAP\ observations.  \cite{bennett/etal:2003c} described
the observations of galactic and extragalactic foreground sources.
\cite{hinshaw/etal:2003} gave the angular power spectrum derived from
the the \iMAP\ maps.  \cite{hinshaw/etal:2003b} described the
\iMAP\ data processing and systematic error limits.
\cite{page/etal:2003b} discussed the beam sizes and window functions
for the \iMAP\ experiment.  \cite{page/etal:2003c} discussed results
that can be derived simply from the positions and heights of the peaks
and valleys in the angular power spectrum.  \cite{spergel/etal:2003}
described the cosmological parameters derived by fitting the
\iMAP\ data and other datasets.  \cite{verde/etal:2003} described the
fitting methods used.  \cite{peiris/etal:2003} described the
consequences of the \iMAP\ results for inflationary models.
\cite{jarosik/etal:2003b} described the on-orbit performance of the
\iMAP\ radiometers.  \cite{kogut/etal:2003} described the
\iMAP\ observations of polarization in the CMB.
\cite{barnes/etal:2003} described the large angle sidelobes of the
\iMAP\ telescopes.  \cite{komatsu/etal:2003} addressed the limits on
non-Gaussianity that can be derived from the \iMAP\ data.

Using the\iMAP\ data plus CBI and ACBAR, the position of the big peak
in the angular power spectrum was found to be $\ell_{pk} = 220.1 \pm
0.8$.  The position of the big peak defines a track in the
$\Omega_m$-$\Omega_V$ plane shown in Figure \ref{fig:Wv-Wm-wSNe-wMAP}.

The ratios of the anisotropy powers below the peak at $\ell \approx
50$, at the big peak at $\ell \approx 220$, in the trough at $\ell
\approx 412$, and at the second peak at $\ell \approx 546$ were
precisely determined using the \iMAP\ data which has a single
consistent calibration for all $\ell$'s.  Previously, these $\ell$
ranges had been measured by different experiments having different
calibrations so the ratios were poorly determined.  Knowing these
ratios determined the photon:baryon:CDM density ratios, and since the
photon density was precisely determined by FIRAS on COBE, accurate
values for the baryon density and the dark matter density were
obtained.  These values are $\Omega_b h^2 = 0.0224 \pm 4\%$, and
$\Omega_m h^2 = 0.135 \pm 7\%$.  The ratio of CDM to baryon densities
from the \iMAP\ data is 5.0:1.

Because the matter density $\Omega_m h^2$ was fairly well constrained
by the amplitudes, the positions of a point in Figure
\ref{fig:Wv-Wm-wSNe-wMAP} served to define a value of the Hubble
constant.  The size of the points in Figure \ref{fig:Wv-Wm-wSNe-wMAP}
indicates how well this derived Hubble constant agrees with the
$H_\circ = 72 \pm 8$ from the HST Key Project
(\cite{freedman/etal:2001}).  Shown as contours are the $\Delta \chi^2
= 1,\;4, \;\&\; 9$ contours from my fits to 230 SNe Ia
(\cite{tonry/etal:2003}).  Clearly the CMB data, the HST data, and the
SNe data are all consistent at a three-way crossing that is very close
to the flat Universe line.  Assuming the Universe actually is flat,
the age of the Universe is very well determined: 
$13.7 \pm 0.2\;\mbox{Gyr}$.

\iMAP\ also found a TE (temperature-polarization) cross-correlation.
At small angles the TE amplitude was perfectly consistent with the
standard picture of the recombination era.  But there was also a large
angle TE signal that gave an estimate for the electron scattering
optical depth since reionization: $\tau = 0.17 \pm 0.04$.  Based on
this the epoch of reionization was 200 million years after the Big Bang.

\section{Summary}

Measurements of the CMB anisotropy made in the last 5 years have
moved cosmology into a new era of precise parameter determination
and the ability to probe the conditions during the inflationary
epoch.  These results depend on the study of perturbations that
are still in the linear, small amplitude regime, and thus are not
confounded by non-linearities and the difficulties associated
with hydrodynamics.

\begin{acknowledgments}
The \iMAP\ mission is made possible by the support of the Office of Space
Sciences at NASA Headquarters and by the hard and capable work of scores of
scientists, engineers, technicians, machinists, data analysts, budget analysts,
managers, administrative staff, and reviewers.
\end{acknowledgments}

\bibliographystyle{kapalike}

\end{document}